\documentclass[preprint,amsmath,amssymb]{revtex4-1}
\usepackage{graphicx}
\usepackage{dcolumn}
\usepackage{bm}
\usepackage{mathrsfs}
\newcommand{\pd}[2]{\frac{\partial #1}{\partial #2}} 
\newcommand{\ket}[1]{\left| #1 \right>} 
\newcommand{\bra}[1]{\left< #1 \right|} 

 \def \figscale{1.5}
\begin{document}

\title{Super-Sensitive Ancilla-Based Adaptive Quantum Phase Estimation}
\author{Walker Larson}
\author{Bahaa Saleh}
 \email{besaleh@creol.ucf.edu}
\affiliation{
CREOL, The College of Optics and Photonics, University of Central Florida, Orlando, Florida 32816, USA}
\date{\today}

\begin{abstract} 
The super-sensitivity attained in quantum phase estimation is known to be compromised in the presence of decoherence.  This is particularly patent at blind spots -- phase values at which sensitivity is totally lost.   One remedy is to use a precisely known reference phase to shift the operation point of the sensor to a less vulnerable phase value.  We present here an alternative approach based on combining the probe with an ancillary degree of freedom containing adjustable parameters to create an entangled quantum state of higher dimension.  We validate this concept by simulating a configuration of a Mach-Zehnder interferometer with a two-photon probe and a polarization ancilla of adjustable parameters, entangled at a polarizing beam splitter. At the interferometer output, the photons are measured after an adjustable unitary transformation in the polarization subspace.  Through calculation of the Fisher information and simulation of an adaptive estimation procedure, we show that optimizing the adjustable polarization parameters using an adaptive measurement process provides globally super-sensitive unbiased phase estimates for a range of decoherence levels, without prior information or a reference phase. 
\end{abstract}

\maketitle

\section{\label{sec:into}Introduction}
It is well-known that bounds on the sensitivity of optical measurements based on coherent-state probes can be surpassed by use of non-classical light \cite{Giovannetti2011,Simon2016,Demkowicz-Dobrzanski2012,Holland1993,Braunstein1995,Banaszek2009}. 
As a canonical example, estimates of the optical phase are bounded by the shot-noise or classical limit (CL) in classical sensing strategies and by the Heisenberg-limit (HL) \cite{Lee2002,Giovannetti2004,Giovannetti2006} in non-classical sensing strategies.  As dictated by the Cram\'{e}r-Rao bound \cite{Helstrom1969,Paris2009}, the variance of estimates employing an average of $N$ photons can scale at best as $\frac{1}{\sqrt{N}}$ for classical probes, while the variance of estimates employing exactly $N$ photons can scale at best as $\frac{1}{N}$; an estimate that achieves a variance between these ranges is commonly referred to as super-sensitive.   

The super-sensitivity attained in quantum phase estimation is compromised in the presence of any finite source of quantum state imperfection or decoherence \cite{Okamoto2008,Shaji2007,Koodynski2013,Dowling2008}, making it rather challenging to reach the HL goal.  Even worse, for certain values of phase, which we refer to as blind spots, the measurement fails to provide \textit{any} sensitivity \cite{Israel2014} \cite{Matthews2016}. Traditional adaptive phase estimation overcomes this issue by employing a reference phase and iteratively moving the operation point of the interferometer to the range for which it is most sensitive. To our knowledge, in every demonstration of the use of two-photon interferometry, a reference phase has been required to observe super-sensitivity. 

Lately, much work has been done to investigate how ancillary photons or degrees of freedom (DoFs) can be used to aid quantum estimation strategies against such deleterious effects \cite{Huang2016,Demkowicz-Dobrzanski2014,Dur2014}.  More specifically, recent work \cite{Jachura2016} has shown that in the presence of partial two-photon spectral distinguishability, an effect that degrades two-photon interference while leaving single-photon interference unhindered, it is possible to employ an ancillary DoF to fortify super-sensitive two-photon states against the total loss of sensitivity at the blind spots. By coupling an ancillary DoF (ancilla) to the probe DoF it was possible to theoretically model and experimentally measure sensitivity above the CL using coincidence measurements at a blind spot. 
		
In the present work, we consider ancilla-based phase estimation with a two-photon quantum state impaired by decoherence described by the 'depolarizing-channel' model, which is one of the most general models of decoherence in two-photon system. This effect degrades both two-photon and single-photon interference.    We use a configuration of a Mach-Zehnder interferometer with a
two-photon probe and a polarization ancilla of adjustable parameters, entangled at a polarizing beam splitter. At the interferometer output, the photons are measured after an adjustable unitary
transformation in the polarization subspace. Through calculation of the Fisher information we show that fortification through an ancillary DoF protects the quantum advantage afforded to two-photon measurements for a range the depolarization probabilities (decoherence levels). Within this range, it is possible to use the ancillary DoF, rather than a reference phase, which  must be placed within the interferometer itself, to retain the sensitivity of the interferometer.

We also show that adaptive phase estimation can be performed in this paradigm by tuning the polarization (ancilla) of the input two-photon states and the two-photon polarization measurements that are made at the output of the system, rather than tuning the optical system itself. Previous experimental and theoretical treatments of this topic have only considered the case where precise prior information of the phase exists, and this is, to our knowledge, the first theoretical work that considers the entire adaptive process. Our simulations suggest that just as in the case of using a reference phase, adaptive tuning of the ancillary degree of freedom provides unbiased estimators that are super-sensitive for a range of decoherence probabilities. The techniques developed here can therefore play a critical role in phase estimation tasks where introduction of a reference phase is not feasible. 

\section{ Effect of Decoherence on Phase Sensitivity}

\subsection{Two-photon probe in a pure state}
The phase $\phi$ introduced by transmission through an optical element is typically measured by placing the element in one arm of a MZI and using an optical probe at the input ports together with an appropriate measurement at the output ports.  An unbiased phase estimate $\tilde{\phi}$ based on a measurement outcome $M$ has a statistical variance satisfying the Cram\'er-Rao  bound, ${\mathrm{Var}}(\tilde{\phi}) \geq  \frac{1}{F^(\phi)}$, where $F(\phi)= p( M |\phi)  \left[ \pd{}{\phi} \ln p( M |\phi) \right]^2$ is the Fisher information  and $p( M |\phi)$ is the conditional probability distribution of measuring $M$ given $\phi$. This variance, which defines the sensitivity of the measurement, clearly depends on the choice of the probe and the measurement.
        
                Here, we limit ourselves to two-photon optical probe states and two-photon measurements --- either through coincidence between single-photon detection, photon-number resolving detection, or both. When the quantum state is a pure state with one photon in each of the interferometer input ports, it turns out that measuring two-photon coincidence and double counts at the output ports of the interferometer is in fact the optimal measurement, with $F(\phi)=4$, $\forall$ $\phi$, achieving the HL that corresponds to the highest sensitivity achievable using two photons.  For comparison, a classical optical probe in a coherent state with an average of two photons obtains a maximum value of $F(\phi)=2$.   In this case, the quantum two-photon probe offers a factor of 2 advantage in the variance of estimates over a classical probe with the same mean number of photons \cite{Kacprowicz2010,Ono2013}.
    
    \subsection{Two-photon probe with decoherence} 
One would expect that the phase sensitivity achievable with an optical probe in a two-photon state subjected to decoherence would deteriorate gradually as the strength of decoherence increases. It turns out that the effect of decoherence also depends significantly on the actual value of the phase $\phi$. To verify this, we subject the otherwise pure input state to the 'depolarization' channel of decoherence. This is a model that randomly replaces the input state describing a single DoF of a photon (here the interferometer path mode) with a  mixed  state. The process occurs with probability $p$. The channel-sum representation of the operation can be used to model depolarization acting on a multi-degree-of-freedom or multi-photon states (See appendix).
	
\begin{figure}[h]
\begin{center}
\includegraphics[scale=\figscale]{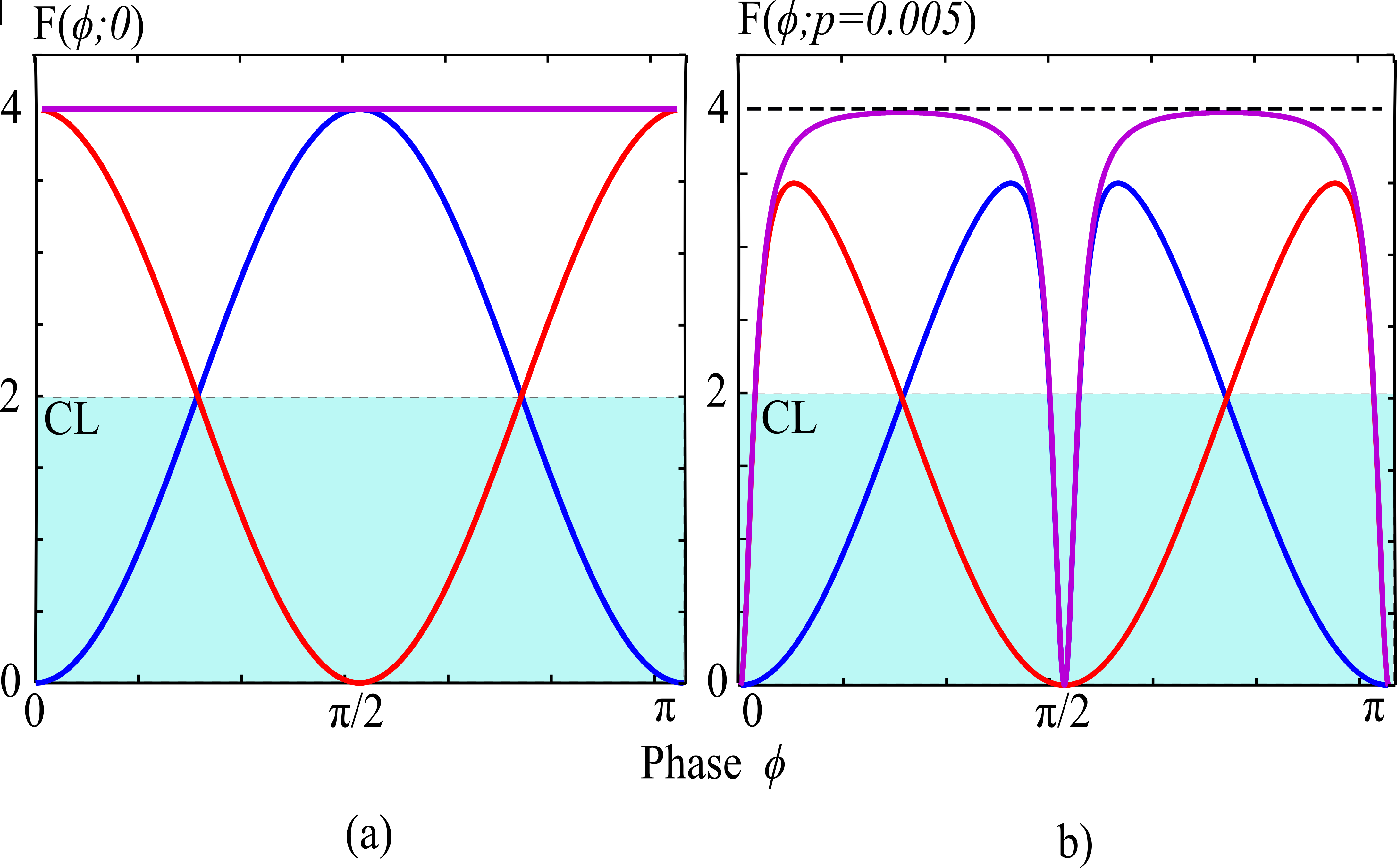}
\caption{Fisher information as a function of the phase $\phi$ for two cases. (a) When a MZI is fed with the pure two-photon input state, the sum of the Fisher information provided by coincidence (blue) and double-count (red) measurements leads to the total (purple) $F(\phi;0)$ that can be gained from two-photon measurements at the output of an ideal Mach-Zehnder interferometer. (b) As in (a), but  with non-zero decoherence probability $p$ (shown for $p=0.005$). Blind spots appear at $\phi=0,\frac{\pi}{2},\pi$ when using the same strategy that achieved the Heisenberg limit in the absence of decoherence.}
\label{DepolMZI}
\end{center}
\end{figure}

	To investigate how this affects the sensitivity of our system, we calculated the Fisher information $F(\phi;p)$ using the depolarized input state. Towards this calculation, a pure input state describing the interferometer-path-modes (probe) of a photon pair was represented as superpositions of vectors of the form $\ket{P_1} \otimes \ket{P_2}$, where $P$ denotes the binary probe DoF, and $1,2$ refer to the first and second photon. To create the optimal probe state \cite{C.K.Hong1987},
 \begin{equation}
\ket{\psi_0}=\tfrac{1}{\sqrt{2}} \left[\ket{u}\ket{l}+\ket{u}\ket{l}\right],
\end{equation}  
where $\ket{u}$ and $\ket{l}$ correspond to the upper and lower input ports of the MZI, respectively, was used. The density matrix $\rho_0=\ket{\psi_0}\bra{\psi_0}$ of this pure state is altered by decoherence, becoming a mixed state 
	\begin{equation}
	\rho(p)= \mathcal{E}_p(  \rho_0 ),
	\end{equation}
 where $\mathcal{E}_p$ represents the depolarization operation, which is a function of  the probability $p$ (see appendix). This state is then evolved unitarily by the interferometer, encoding information about the phase difference $\phi$ into the state, leaving the output state 
	\begin{equation}
	\rho(\phi;p)= U(\phi) \rho(p) U^{\dagger} (\phi).
	\end{equation}
The probability of measuring a coincidence count is then
\begin{equation}
	P_c(\phi;p)=\textrm{Tr}[\rho(\phi;p)\Pi_c],
	\end{equation}
 and the probability of measuring a double count is 
 \begin{equation}
 	P_d(\phi;p)=\textrm{Tr}[\rho(\phi;p)\Pi_d],
 \end{equation}
 where
 
 	 \begin{equation}
\Pi_c=\ket{u}\ket{l}\bra{u}\bra{l}+\ket{l}\ket{u}\bra{l}\bra{u},
	 \end{equation}
    and
         
      	 \begin{equation}
\Pi_d=\ket{u}\ket{u}\bra{u}\bra{u}+\ket{l}\ket{l}\bra{l}\bra{l},
	 \end{equation}
    are the  operators describing coincidence and double counts, respectively. 
From these probabilities,  we find  
 	\begin{equation}
	F(\phi;p)=\frac{8 (p-1)^4 \sin^2(2 \phi)}{1-(p-1)^4 \cos(4 \phi)-(p-2) p ((p-2) p+2))},
	\end{equation} which is plotted in Fig \ref{DepolMZI}.

	When the depolarization operation contaminates the input state with any non-zero probability $p$, there is a drastic change in $F(\phi;p)$, and we see the emergence of blind spots --- phases for which the sensitivity afforded by both two-photon measurements drops sharply to zero.  Specifically, for $\phi \in \lbrace \phi_{BS} \rbrace= \left \lbrace 0,\frac{\pi}{2},\pi \right \rbrace$, we find that $F(\phi;p)=0$.  This behavior is plotted in Figure \ref{DepolMZI}. With this drastic change, it is clear that the once-optimal interferometer will now be completely insensitive to phase values in the neighborhood of any of these blind spots. At these spots, finding an unbiased estimator of $\phi$ becomes impossible as $F(\phi;p)$ approaches zero. 
    
\subsection{Reference Phase}
As a simple remedy to the loss of sensitivity at a blind spot, one could add a precisely known, tunable reference phase $\phi_r$ to a path of the interferometer \cite{Ono2013}. To show this, we calculate the Fisher information $F(\phi-\phi_r;p)$ that results from measuring the probabilities of coincidence  and double counts when a reference phase is used. The probability of measuring coincidence or double counts are given by $P_c(\phi-\phi_r;p)=\textrm{Tr}[\rho(\phi-\phi_r;p)\Pi_c]$ or $ 	P_d(\phi-\phi_r;p)=\textrm{Tr}[\rho(\phi-\phi_r;p)\Pi_d]$, respectively, and give a sensitivity described by $F(\phi -\phi_r;p)$.

The optimal sensitivity afforded to this strategy is then given by maximizing $F(\phi -\phi_r;p)$
over $\phi_r$, for which the optimal operating point will be found at $\phi-\phi_r=\frac{\pi}{4}$. This optimization provides
\begin{equation}
F_{\phi_r}(p)=4 (1-p)^4,
\end{equation}
and shows that this strategy retains super-sensitivity for values of $p<0.1591$. For reference, this corresponds to an interference visibility of $\frac{2}{3}$ \cite{Rarity1990} if the depolarization operation is the only source of imperfect visibility.

While the introduction of a phase reference conveniently obviates the repercussions of the phase dependence on sensitivity, we are left with a  question: could we perform some other modification to our system that does not require  physical changes inside the interferometer itself? To answer this question affirmatively, we introduce a second degree of freedom, an ancilla, which we implement by means of the polarization of the photon pair. 

\section{Ancilla Fortification}
\begin{figure}[h]
\begin{center}
\includegraphics[scale=\figscale]{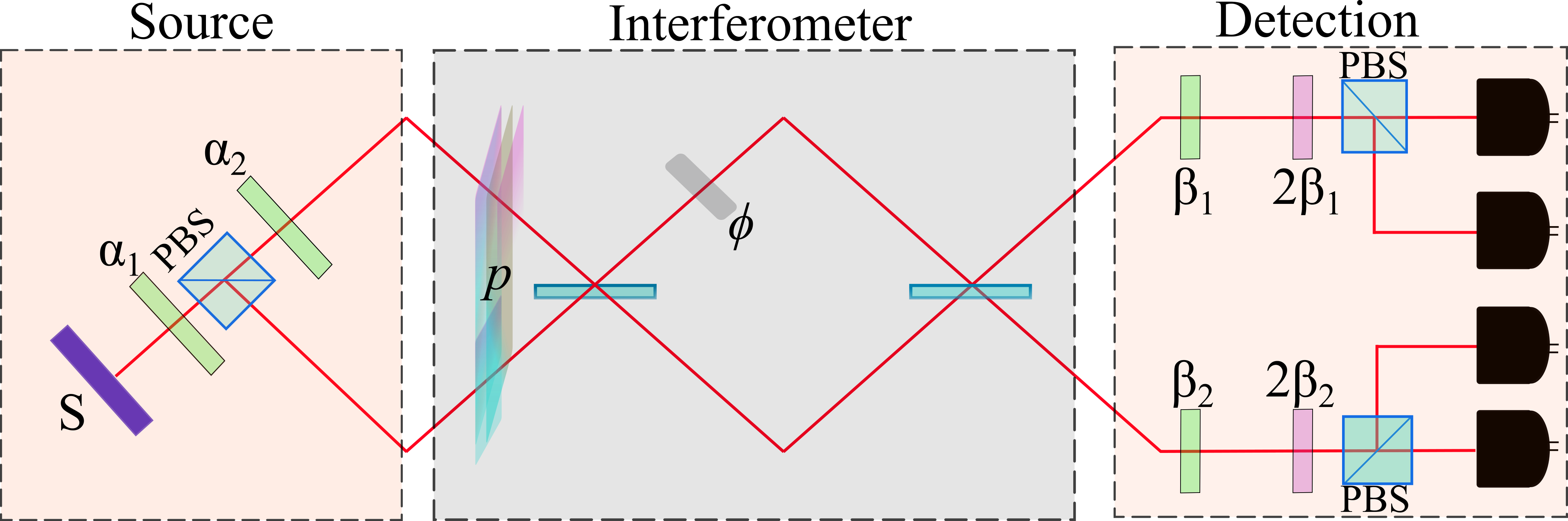}
\caption{A path-polarization two-photon entangled state is created by transmitting photon-pairs generated by a source S in the state $\ket{H}\ket{V}$ through a half-wave plate (HWP) with optic axis at an angle $\alpha_1$, separating the polarization components with a polarizing beam splitter (PBS), and passing one component through a second HWP at angle $\alpha_2$. The entangled state created is subjected to the depolarizing channel as it enters an interferometer with the measured phase $\phi$ in one arm.  At each of the output ports of the interferometer, polarization measurements are tuned by placing a HWP at angle $\beta_{1,2}$, a quarter-wave plate at angle $2\beta_{1,2}$, and a PBS with two detectors at its output are used for photon detection. The angles $\beta_1$ and $\beta_2$ are independent.}
\label{AncillaD}
\end{center}
\end{figure}

In a configuration using polarization  of the photon pair used as an ancillary DoF, the input state begins as a pair of photons with crossed polarizations in a single spatial mode. The pure input state describing the  interferometer-path-modes (probe) and polarization (ancillary) DoFs of a photon pair is represented as linear combinations of vectors of the product form $ \ket{P_1}\ket{A_1} \otimes \ket{P_2}\ket{A_2}$, where $P,A$ denote the binary probe and ancillary DoFs, respectively, and $1,2$ refer to the first and second photon, respectively.  The added DoF creates a large set of possible states; while the ones we will proceed to use are not the most general forms of states in this set, we found them to provide the best sensitivity out of all the combinations that we have studied.
 
We start with an input state with a horizontally polarized photon in the same spatial mode as a vertically polarized photon is subjected to the transformation dictated by transmission through a half-wave plate (HWP) with an optic axis at an angle $\alpha_1$. Spatial-mode-polarization correlations are then created by use of a polarizing beam splitter (PBS)that enacts the transformation $\ket{H} \rightarrow \ket{l}\ket{H} $ and $\ket{V}  \rightarrow \ket{u}\ket{V} $. Correlations are further tuned by placing a second HWP with an optic axis at an angle $\alpha_2$ in the upper arm, after the PBS, as shown in Fig \ref{AncillaD}. The states created by this process are states that, in general, are not optimal when $p=0$. This may not be surprising; for a large number of estimation tasks, the optimal quantum probe states are rarely optimal once  decoherence is introduced to the system \cite{Demkowicz-Dobrzanski2009,Dorner2009,MacCone2009,Escher2011}. The input state after preparation is given by 

  \begin{multline}
 \ket{\psi_{in}}^{(A)}=\cos^2 \alpha_1 \ket{l}\ket{H}\ket{l}\ket{H} -\sin^2 \alpha_1 \ket{u}\ket{\alpha_2}\ket{u}\ket{\alpha_2}\\
 +\cos \alpha_1 \sin \alpha_1 \left( \ket{l}\ket{H}\ket{u}\ket{\alpha_2}- \ket{u}\ket{\alpha_2}\ket{l}\ket{H}\right),
\end{multline}
where the polarization state $\ket{ \alpha_2}=\cos \alpha_2 \ket{H} -\sin \alpha_2 \ket{V}$.

This state is then acted on by the depolarizing channel and transformed by transmission through the interferometer. A pair of  HWPs and quarter-wave plates (QWP) are placed into each of the output arms of the interferometer. The HWPs have optics axes at angles $\beta_1$ (upper arm) and $\beta_2$ (lower arm), while the QWPs have optic axes at angles  $2 \beta_1$ (upper arm)  and $2 \beta_2$ (lower arm). Two-photon measurements are then made at the output ports of a PBS placed after each pair of wave plates. As a result, the output state before the final pair of of PBSs, $\rho^{(a)} (\phi; p, \Theta)$, is completely characterized by the parameter set $\Theta=\left \lbrace \alpha_1, \alpha_2, \beta_1, \beta_2\right \rbrace$. 
 
Now, the calculation of the optimal Fisher information for a given value of $\phi$, $F_{\textrm{opt}}^{(a)}(\phi;p)$, becomes an optimization over the  parameter set $\Theta$ as opposed to a reference phase $\phi_r$. The probabilities of measuring the two photon state  in output path modes $\left\lbrace k_1, k_2\right\rbrace$ and  polarization modes $\lbrace s_1,s_2 \rbrace$ needed to calculate $F_{\textrm{opt}}^{(a)}(\phi; p)$ are given by 
\begin{equation}
P_{k_1,s_1,k_2,s_2}=\textrm{Tr} [ \rho^{(a)} (\phi; p, \Theta) \Pi_{k_1,s_1,k_2,s_2}],
\end{equation}
where
\begin{equation}
\Pi_{k_1,s_1,k_2,s_2}=\ket{k_1}\ket{s_1}\ket{k_2}\ket{s_2}\bra{k_1}\bra{s_1}\bra{k_2}\bra{s_2}.
\end{equation}

For each value of $\phi$, an exhaustive search  over the sensitivity that results from a given set $\Theta$ determines the values that provide the optimal sensitivity $F_{\textrm{opt}}^{(a)}(\phi;p)=\underset{\left\lbrace \Theta \right\rbrace}{\textrm{max}} \left\lbrace F^{(a)}(\phi_;p,\Theta) \right\rbrace$. The sensitivity that results from this optimization is plotted in Figure 3 for $p=0.05$. While operation at the blind spots may not be as sensitive as the operation that could be attained using a reference phase, it is clear that it makes possible the quantum super-sensitive advantage in a sensing regime where no sensitivity was possible prior. We find that super-sensitivity is globally attainable for all phases for decoherence probabilities less than $p=0.072$.

\begin{figure}[h]
\begin{center}
\includegraphics[scale=\figscale]{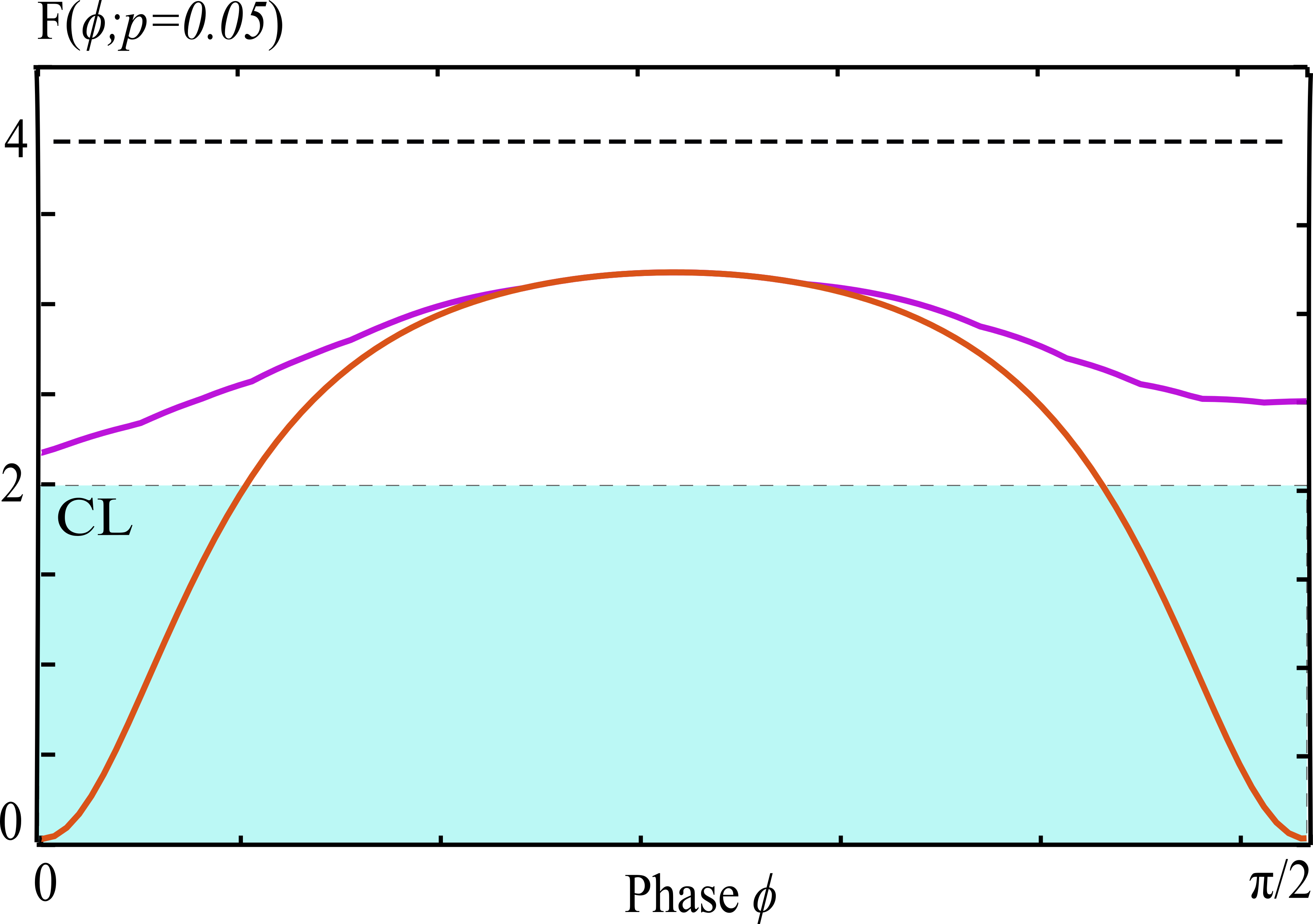}
\caption{Fisher information is plotted for a two-photon interferometer with (purple) and without (orange) the employment of an ancillary degree of freedom. Ancilla fortification allows for measurements that retain sensitivity at blind spots $0$,$\frac{\pi}{2}$. Super-sensitivity is retained for $p < 0.072$ (plotted here for $p=0.05$) when an ancilla is used.  }
\label{RefVAnc}
\end{center}
\end{figure}

 \section{Adaptive Phase Estimation}
 To show that this methodology can find application in the general setting of quantum phase estimation, we have conducted simulations demonstrating how the ancilla-aided strategy provides a full platform for super-sensitive adaptive phase estimation.  Simulations of adaptive phase estimation were conducted by supplementing simulations of maximum likelihood estimation with feedback based on measurement results. Traditional adaptive phase estimations uses feedback from measurements  to update the value of the reference phase $\phi_r$, setting it to the value that maximizes the sensitivity of the optical system, assuming that the true value of $\phi$ is equal to the most recent estimate $\tilde{\phi}$ \cite{Pope2004,Hou2016,Lerch2014,Okamoto2012,Higgins2007}. Likewise, when using ancilla fortified states without a reference phase, feedback updates the parameter set $\Theta$,  setting the parameters equal to the set that maximize the sensitivity of the system for the assumed value of $\phi=\tilde{\phi}$.
 
 Specifically, performing an $n^{th}$ two-photon measurement provides a  result $M_n$ that corresponds to the measurement operator element  $\Pi_n \in \Pi_{k_1,k_2,s_1,s_2}$. The likelihood of the result is $\mathcal{L}_n (\phi)=\textrm{Tr}\left[\rho(\phi) \Pi_n\right] \mathcal{L}_{n-1} (\phi)$, from which an $n^{th}$ estimate of $\phi$, $\tilde{\phi}_n=\textrm{argmax}_{\phi}\textrm{ }\mathcal{L}_n (\phi)$, is made.  With the estimate $\tilde{\phi}_n$, the settings of the parameters $\Theta$ that maximize  $F^{(a)}(\tilde{\phi}_n;p,\Theta)$ are updated \cite{Chapeau-Blondeau2016}. To ensure that final estimate $\tilde{\phi}_N$ after $N$ measurements is unbiased (i.e. $\left \langle \tilde{\phi}_N\right \rangle=\phi$), the initial estimate $\tilde{\phi_0}$ is chosen at random. Finally, to avoid degeneracy in the final likelihood function (which would lead to equally likely estimates spaced some period apart), a final, non-optimal series of measurements must be made. By experimentation, we have found that setting $\Theta$ to the values that are optimal at $\phi=0$, then $\phi=\frac{\pi}{2}$, for the last $3 \%$ of adaptive iterations is sufficient in avoiding this discrepancy. 
 
\begin{figure}[ht]
\begin{center}
\includegraphics[scale=\figscale]{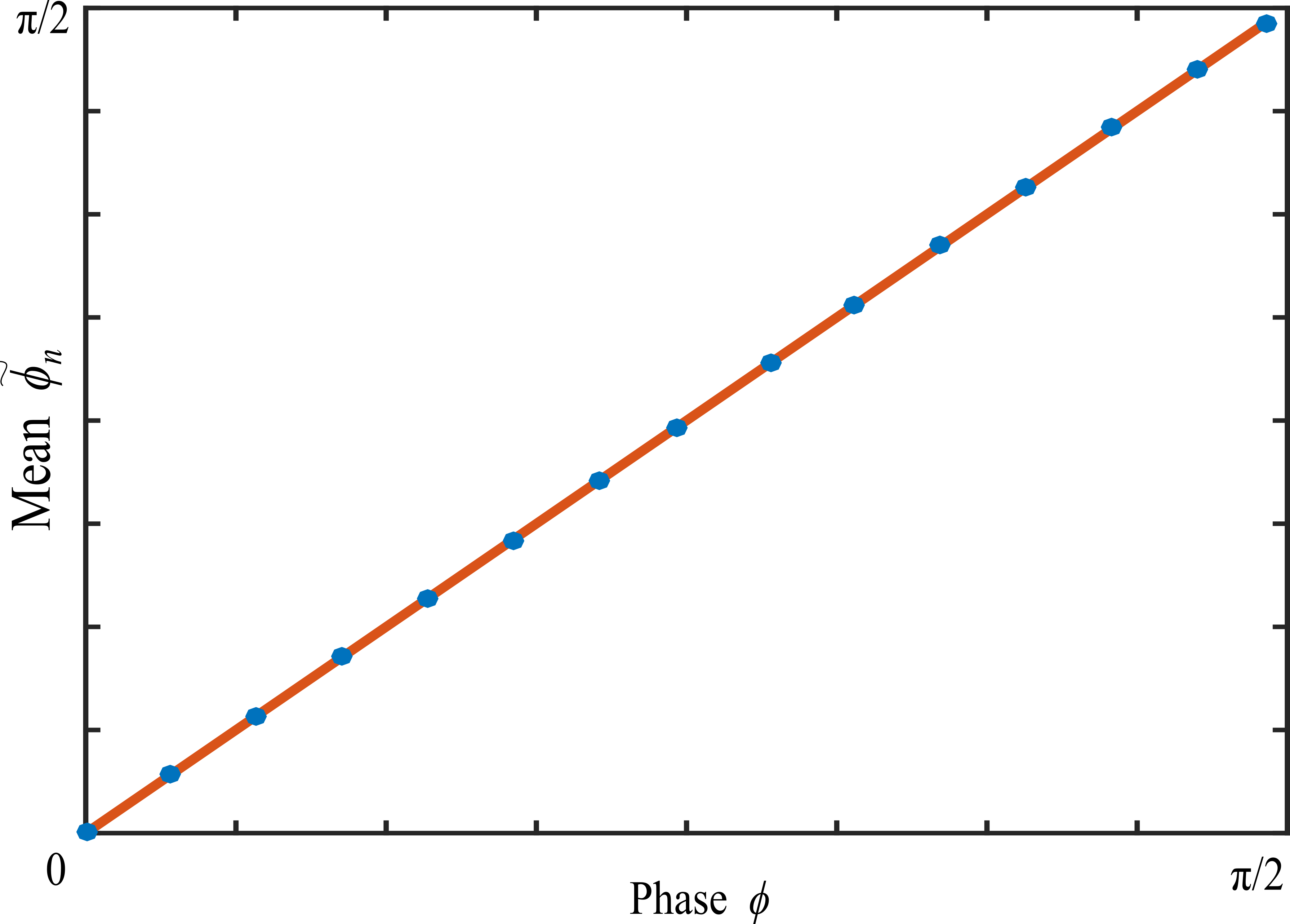}
\caption{Mean of final phase estimates after $S=5000$ simulations of $N=1500$ adaptive two-photon state detections (Blue) are plotted for $p=0.01$ with the 1:1 correspondence expected for unbiased estimators (Orange). }
\label{Mean}
\end{center}
\end{figure}

\begin{figure}[ht]
\begin{center}
\includegraphics[scale=\figscale]{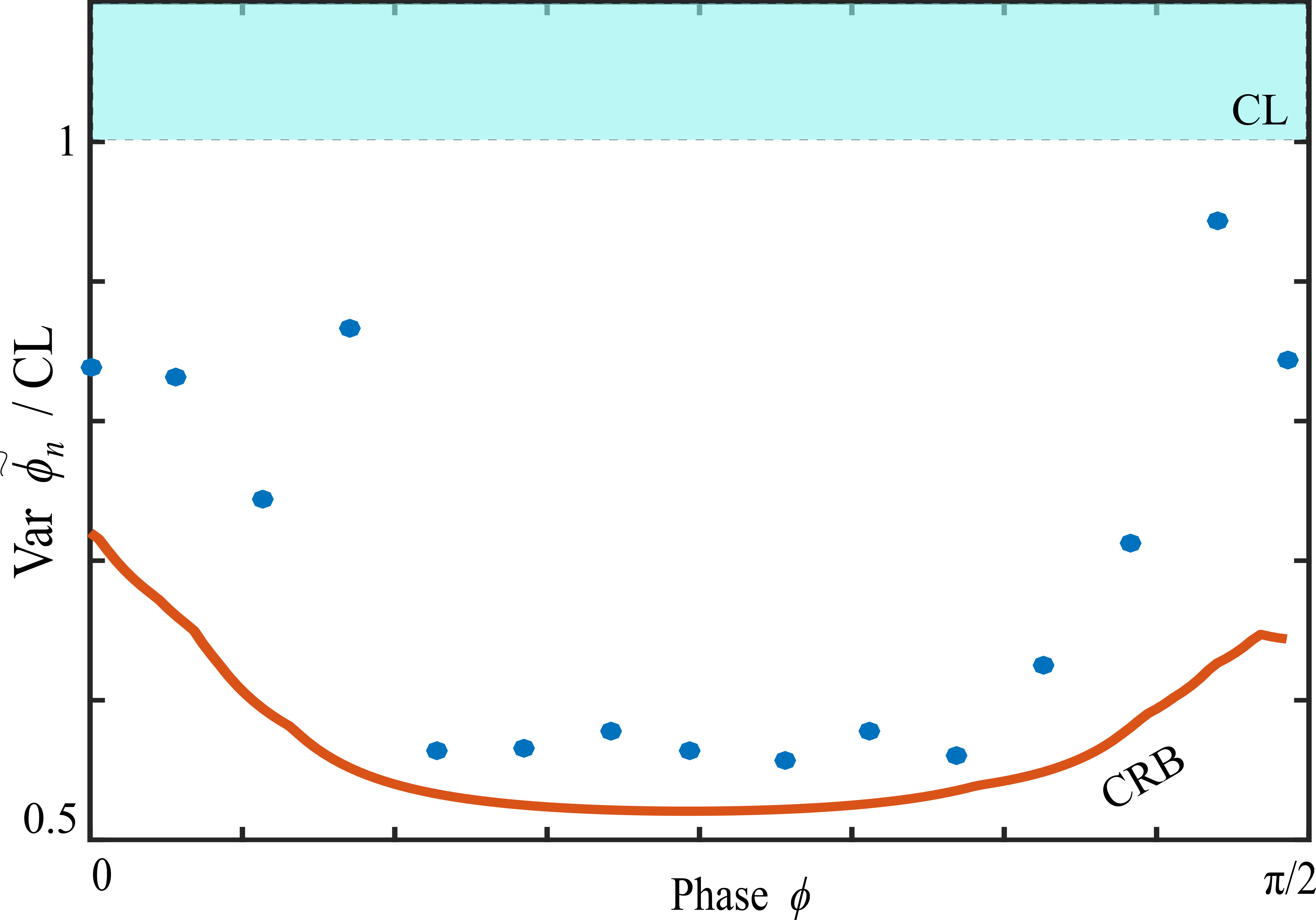}
\caption{Variance of final phase estimates after $S=5000$ simulations of $N=1500$ adaptive two-photon state detections (Blue)  for $p=0.01$ follows the functional form of the Cram\'er-Rao Bound as dictated by the Fisher information (Orange). For each value of $\phi$, a variance below the classical estimation limit is observed.}
\label{Variance}
\end{center}
\end{figure}

For our testing of this procedure for each phase $\phi$, we set $p=0.01$ and performed $S=5000$ simulations, each detecting adaptively a total of $N=1500$ two-photon states. From these simulated trials,  $S$ final estimates $\tilde{\phi}_N$ were collected, and statistics were calculated on the ensemble of these estimates. The mean and variance of estimates as a function of the true value of $\phi$ for each simulation is plotted in Figures \ref{Mean} and \ref{Variance}. For the given sample size, we find that the strategy is unbiased, and the functional dependence of the variance on $\phi$ generally follows the trend predicted by the calculation of the Fisher information.  Most importantly, for this probability of decoherence, our simulations show estimates that are super-sensitive for all values of $\phi$.
 
\section{Conclusion}
 
 The accuracy of a metrological measurement is limited by inherent noise in the probe.  While quantum optical probes offer a global advantage over classical probes, they can be more vulnerable to minute imperfections or contamination by weak extraneous noise.  A case in point is the interferometric measurement of phase by use of a two-photon probe. Although this quantum probe offers a sensitivity advantage (super-sensitivity) under ideal conditions, if the probe is subjected to weak decoherence, the sensitivity will be globally reduced and -- surprisingly -- lost altogether at certain phase values (blind spots).  While the quantum advantage may be partially regained by shifting the phase to a less vulnerable value, the insertion of a precisely known reference phase into the interferometer may not be feasible.  
 
We have investigated an alternative approach based on supplementing the path DoF of the probe with an ancillary DoF (polarization), and creating an entangled quantum state in a Hilbert space of twice the dimensionality.  We have demonstrated that the diversity added into the probe can help avoid the blind-spot predicament. This of course requires tweaking polarization parameters of unitary transformations at both the input and output ports of the interferometer before detecting the outgoing two photons.  These transformations tailor the input state and the corresponding output detection strategy for optimal estimation. In this paper, the number of these adjustable parameters was limited to four, two at the input of the interferometer and two at the output.  Since values of the parameters that maximize the sensitivity depend on the unknown phase itself, the optimization must be conducted adaptively.   Based on extensive simulation of the adaptive process, we conclude that for a range of decoherence strengths, super-sensitivity is indeed obtained for any phase.
 
While the example investigated in this paper uses path and polarization as the principal and ancillary DoFs, the reverse is possible and other DoFs, may also be used, as long as implementation of the prerequisite unitary transformations are practical. Likewise, we expect this methodology to find application in the larger of optical parameter estimation. In the more general task of estimating a unitary transformation, it is expected that blind spots will appear in the estimation of any of the parameters that encode the transformation: in this case, a binary DoF offering a secondary channel for enhanced estimation may be the only way to overcome these parameter blind spots.
\section{Appendix}
Decoherence of a quantum state is described by a transformation that converts the density operator $\rho$ into a density operator 
\begin{equation}
\mathcal{E}(\rho)=\sum_{i} E_i \rho E_i^{\dagger},
\end{equation}
where $E_i$ are appropriate operation elements of this operator-sum representation. 

In the case of the depolarizing channel acting on a binary quantum state, the operation elements are given by 
\begin{equation}
\begin{aligned}
E_1=\sqrt{1-\frac{3p}{4}}\mathcal{I}, \qquad
E_2=\sqrt{\frac{p}{4}}\sigma_x, \\
E_3=\sqrt{\frac{p}{4}}\sigma_y,  \qquad
E_4=\sqrt{\frac{p}{4}}\sigma_z,
\end{aligned}
\end{equation}
where $\hat{\sigma_i}$ are the two-dimensional Pauli matrices, and $\mathcal{I}$ is the two-dimensional identity matrix \cite{Nielsen2011}.

For composite systems, the decoherence operation can apply to individual subsystems. For example if $\rho=\rho^A \otimes \rho^B$, decoherence acting on subsystem $A$ alone is implemented by the operation elements

\begin{equation}
E_i^{AB}=E_i^A \otimes \mathcal{I}^B,
\end{equation}
where superscripts $A$ and $B$ denote transformations acting on respective systems $A$ and $B$. 
With this definition, decoherence can act independently on any combination of degrees of freedom  (DoF) of either photon.

In this paper, we consider a two-photon state with two DoFs, path and polarization, with polarization playing the role of the ancilla.  Decoherence acts equally on both photons in the path DoF, leaving the polarization DoF unaltered. The operation is 
	\begin{equation}
	\mathcal{E}_p(\rho)= \mathcal{E}^{(1)}_p( \mathcal{E}^{(2)}_p( \rho ))=\mathcal{E}^{(2)}_p( \mathcal{E}^{(1)}_p( \rho )),
	\end{equation}
where $\mathcal{E}^{(i)}_p$  denotes the decoherence channel acting on the interferometer-path degree of freedom of the $i^{th}$ photon with probability $p$.


\begin{thebibliography}{36}%
\makeatletter
\providecommand \@ifxundefined [1]{%
 \@ifx{#1\undefined}
}%
\providecommand \@ifnum [1]{%
 \ifnum #1\expandafter \@firstoftwo
 \else \expandafter \@secondoftwo
 \fi
}%
\providecommand \@ifx [1]{%
 \ifx #1\expandafter \@firstoftwo
 \else \expandafter \@secondoftwo
 \fi
}%
\providecommand \natexlab [1]{#1}%
\providecommand \enquote  [1]{``#1''}%
\providecommand \bibnamefont  [1]{#1}%
\providecommand \bibfnamefont [1]{#1}%
\providecommand \citenamefont [1]{#1}%
\providecommand \href@noop [0]{\@secondoftwo}%
\providecommand \href [0]{\begingroup \@sanitize@url \@href}%
\providecommand \@href[1]{\@@startlink{#1}\@@href}%
\providecommand \@@href[1]{\endgroup#1\@@endlink}%
\providecommand \@sanitize@url [0]{\catcode `\\12\catcode `\$12\catcode
  `\&12\catcode `\#12\catcode `\^12\catcode `\_12\catcode `\%12\relax}%
\providecommand \@@startlink[1]{}%
\providecommand \@@endlink[0]{}%
\providecommand \url  [0]{\begingroup\@sanitize@url \@url }%
\providecommand \@url [1]{\endgroup\@href {#1}{\urlprefix }}%
\providecommand \urlprefix  [0]{URL }%
\providecommand \Eprint [0]{\href }%
\providecommand \doibase [0]{http://dx.doi.org/}%
\providecommand \selectlanguage [0]{\@gobble}%
\providecommand \bibinfo  [0]{\@secondoftwo}%
\providecommand \bibfield  [0]{\@secondoftwo}%
\providecommand \translation [1]{[#1]}%
\providecommand \BibitemOpen [0]{}%
\providecommand \bibitemStop [0]{}%
\providecommand \bibitemNoStop [0]{.\EOS\space}%
\providecommand \EOS [0]{\spacefactor3000\relax}%
\providecommand \BibitemShut  [1]{\csname bibitem#1\endcsname}%
\let\auto@bib@innerbib\@empty
\bibitem [{\citenamefont {Giovannetti}\ \emph {et~al.}(2011)\citenamefont
  {Giovannetti}, \citenamefont {Lloyd},\ and\ \citenamefont
  {Maccone}}]{Giovannetti2011}%
  \BibitemOpen
  \bibfield  {author} {\bibinfo {author} {\bibfnamefont {V.}~\bibnamefont
  {Giovannetti}}, \bibinfo {author} {\bibfnamefont {S.}~\bibnamefont {Lloyd}},
  \ and\ \bibinfo {author} {\bibfnamefont {L.}~\bibnamefont {Maccone}},\ }\href
  {\doibase 10.1038/nphoton.2011.35} {\bibfield  {journal} {\bibinfo  {journal}
  {Nature Photonics}\ }\textbf {\bibinfo {volume} {5}},\ \bibinfo {pages} {222}
  (\bibinfo {year} {2011})} \BibitemShut {NoStop}%
\bibitem [{\citenamefont {Simon}(2016)}]{Simon2016}%
  \BibitemOpen
  \bibfield  {author} {\bibinfo {author} {\bibfnamefont {D.~S.}\ \bibnamefont
  {Simon}},\ }\href {\doibase 10.1155/2016/6051286} {\bibfield  {journal}
  {\bibinfo  {journal} {Journal of Sensors}\ }\textbf {\bibinfo {volume}
  {2016}} \bibinfo {pages} {6051286} (\bibinfo {year} {2016})}\BibitemShut {NoStop}%
\bibitem [{\citenamefont {Demkowicz-Dobrza{\'{n}}ski}\ \emph
  {et~al.}(2012)\citenamefont {Demkowicz-Dobrza{\'{n}}ski}, \citenamefont
  {Ko{\l}ody{\'{n}}ski},\ and\ \citenamefont
  {Guţă}}]{Demkowicz-Dobrzanski2012}%
  \BibitemOpen
  \bibfield  {author} {\bibinfo {author} {\bibfnamefont {R.}~\bibnamefont
  {Demkowicz-Dobrza{\'{n}}ski}}, \bibinfo {author} {\bibfnamefont
  {J.}~\bibnamefont {Ko{\l}ody{\'{n}}ski}}, \ and\ \bibinfo {author}
  {\bibfnamefont {M.}~\bibnamefont {Guţă}},\ }\href {\doibase
  10.1038/ncomms2067} {\bibfield  {journal} {\bibinfo  {journal} {Nature
  Communications}\ }\textbf {\bibinfo {volume} {3}} \bibinfo {pages} {1063}
  (\bibinfo {year} {2012})} \BibitemShut {NoStop}%
\bibitem [{\citenamefont {Holland}\ and\ \citenamefont
  {Burnett}(1993)}]{Holland1993}%
  \BibitemOpen
  \bibfield  {author} {\bibinfo {author} {\bibfnamefont {M.~J.}\ \bibnamefont
  {Holland}}\ and\ \bibinfo {author} {\bibfnamefont {K.}~\bibnamefont
  {Burnett}},\ }\href {\doibase 10.1103/PhysRevLett.71.1355} {\bibfield
  {journal} {\bibinfo  {journal} {Physical Review Letters}\ }\textbf {\bibinfo
  {volume} {71}},\ \bibinfo {pages} {1355} (\bibinfo {year}
  {1993})}\BibitemShut {NoStop}%
\bibitem [{\citenamefont {Braunstein}\ \emph {et~al.}(1995)\citenamefont
  {Braunstein}, \citenamefont {Caves},\ and\ \citenamefont
  {Milburn}}]{Braunstein1995}%
  \BibitemOpen
  \bibfield  {author} {\bibinfo {author} {\bibfnamefont {S.~L.}\ \bibnamefont
  {Braunstein}}, \bibinfo {author} {\bibfnamefont {C.~M.}\ \bibnamefont
  {Caves}}, \ and\ \bibinfo {author} {\bibfnamefont {G.~J.}\ \bibnamefont
  {Milburn}},\ }\href {\doibase 10.1006/aphy.1996.0040} {\bibfield  {journal}
  {\bibinfo  {journal} {Annals of Physics}\ }\textbf {\bibinfo {volume}
  {173}}\bibinfo {pages} {39} (\bibinfo {year} {1995})},\ \BibitemShut
  {NoStop}%
\bibitem [{\citenamefont {Banaszek}\ \emph {et~al.}(2009)\citenamefont
  {Banaszek}, \citenamefont {Demkowicz-Dobrzanski},\ and\ \citenamefont
  {Walmsley}}]{Banaszek2009}%
  \BibitemOpen
  \bibfield  {author} {\bibinfo {author} {\bibfnamefont {K.}~\bibnamefont
  {Banaszek}}, \bibinfo {author} {\bibfnamefont {R.}~\bibnamefont
  {Demkowicz-Dobrzanski}}, \ and\ \bibinfo {author} {\bibfnamefont {I.~A.}\
  \bibnamefont {Walmsley}},\ }\href {\doibase 10.1038/nphoton.2009.223}
  {\bibfield  {journal} {\bibinfo  {journal} {Nature Photonics}\ }\textbf
  {\bibinfo {volume} {3}},\ \bibinfo {pages} {673} (\bibinfo {year} {2009})} \BibitemShut
  {NoStop}%
\bibitem [{\citenamefont {Lee}\ \emph {et~al.}(2002)\citenamefont {Lee},
  \citenamefont {Kok},\ and\ \citenamefont {Dowling}}]{Lee2002}%
  \BibitemOpen
  \bibfield  {author} {\bibinfo {author} {\bibfnamefont {H.}~\bibnamefont
  {Lee}}, \bibinfo {author} {\bibfnamefont {P.}~\bibnamefont {Kok}}, \ and\
  \bibinfo {author} {\bibfnamefont {J.~P.}\ \bibnamefont {Dowling}},\ }\href
  {\doibase 10.1080/0950034021000011536} {\bibfield  {journal} {\bibinfo
  {journal} {Journal of Modern Optics}\ }\textbf {\bibinfo {volume} {49}},\
  \bibinfo {pages} {2325} (\bibinfo {year} {2002})}\BibitemShut
  {NoStop}%
\bibitem [{\citenamefont {Giovannetti}\ \emph {et~al.}(2004)\citenamefont
  {Giovannetti}, \citenamefont {Lloyd},\ and\ \citenamefont
  {Maccone}}]{Giovannetti2004}%
  \BibitemOpen
  \bibfield  {author} {\bibinfo {author} {\bibfnamefont {V.}~\bibnamefont
  {Giovannetti}}, \bibinfo {author} {\bibfnamefont {S.}~\bibnamefont {Lloyd}},
  \ and\ \bibinfo {author} {\bibfnamefont {L.}~\bibnamefont {Maccone}},\ }\href
  {\doibase 10.1126/science.1104149} {\bibfield  {journal} {\bibinfo  {journal}
  {Science}\ }\textbf {\bibinfo {volume} {306}},\ \bibinfo {pages} {1330}
  (\bibinfo {year} {2004})}\BibitemShut {NoStop}%
\bibitem [{\citenamefont {Giovannetti}\ \emph {et~al.}(2006)\citenamefont
  {Giovannetti}, \citenamefont {Lloyd},\ and\ \citenamefont
  {Maccone}}]{Giovannetti2006}%
  \BibitemOpen
  \bibfield  {author} {\bibinfo {author} {\bibfnamefont {V.}~\bibnamefont
  {Giovannetti}}, \bibinfo {author} {\bibfnamefont {S.}~\bibnamefont {Lloyd}},
  \ and\ \bibinfo {author} {\bibfnamefont {L.}~\bibnamefont {Maccone}},\ }\href
  {\doibase 10.1103/PhysRevLett.96.010401} {\bibfield  {journal} {\bibinfo
  {journal} {Physical Review Letters}\ }\textbf {\bibinfo {volume} {96}},\
  \bibinfo {pages} {010401} (\bibinfo {year} {2006})}
  \BibitemShut {NoStop}%
\bibitem [{\citenamefont {Helstrom}(1969)}]{Helstrom1969}%
  \BibitemOpen
    \bibfield  {author} {\bibinfo {author} {\bibfnamefont {C.~W.}\ \bibnamefont
  {Helstrom}},\ }\href {\doibase 10.1007/BF01007479} {\emph {\bibinfo
  {Quantum Detection and Estimation Theory} {{Academic Press}},}\ } (\bibinfo {year}
  {1976})\BibitemShut {NoStop}%
\bibitem [{\citenamefont {Paris}(2009)}]{Paris2009}%
  \BibitemOpen
  \bibfield  {author} {\bibinfo {author} {\bibfnamefont {M.~G.~A.}\
  \bibnamefont {Paris}},\ }\href {\doibase 10.1142/S0219749909004839}
  {\bibfield  {journal} {\bibinfo  {journal} {International Journal of Quantum
  Information}\ }\textbf {\bibinfo {volume} {07}},\ \bibinfo {pages} {125}
  (\bibinfo {year} {2009})}\BibitemShut {NoStop}%
\bibitem [{\citenamefont {Okamoto}\ \emph {et~al.}(2008)\citenamefont
  {Okamoto}, \citenamefont {Hofmann}, \citenamefont {Nagata}, \citenamefont
  {O'Brien}, \citenamefont {Sasaki},\ and\ \citenamefont
  {Takeuchi}}]{Okamoto2008}%
  \BibitemOpen
  \bibfield  {author} {\bibinfo {author} {\bibfnamefont {R.}~\bibnamefont
  {Okamoto}}, \bibinfo {author} {\bibfnamefont {H.~F.}\ \bibnamefont
  {Hofmann}}, \bibinfo {author} {\bibfnamefont {T.}~\bibnamefont {Nagata}},
  \bibinfo {author} {\bibfnamefont {J.~L.}\ \bibnamefont {O'Brien}}, \bibinfo
  {author} {\bibfnamefont {K.}~\bibnamefont {Sasaki}}, \ and\ \bibinfo {author}
  {\bibfnamefont {S.}~\bibnamefont {Takeuchi}},\ }\href {\doibase
  10.1088/1367-2630/10/7/073033} {\bibfield  {journal} {\bibinfo  {journal}
  {New Journal of Physics}\ }\textbf {\bibinfo {volume} {10}} \bibinfo{pages}{073033} (\bibinfo {year}
  {2008})}\BibitemShut {NoStop}%
\bibitem [{\citenamefont {Shaji}\ and\ \citenamefont
  {Caves}(2007)}]{Shaji2007}%
  \BibitemOpen
  \bibfield  {author} {\bibinfo {author} {\bibfnamefont {A.}~\bibnamefont
  {Shaji}}\ and\ \bibinfo {author} {\bibfnamefont {C.~M.}\ \bibnamefont
  {Caves}},\ }\href {\doibase 10.1103/PhysRevA.76.032111} {\bibfield  {journal}
  {\bibinfo  {journal} {Physical Review A}\ }\textbf {\bibinfo {volume} {76}},\ \bibinfo {pages} {1} (\bibinfo
  {year} {2007})} \BibitemShut {NoStop}%
\bibitem [{\citenamefont {Ko{\l}ody{\'{n}}ski}\ and\ \citenamefont
  {Demkowicz-Dobrzanski}(2013)}]{Koodynski2013}%
  \BibitemOpen
  \bibfield  {author} {\bibinfo {author} {\bibfnamefont {J.}~\bibnamefont
  {Ko{\l}ody{\'{n}}ski}}\ and\ \bibinfo {author} {\bibfnamefont
  {R.}~\bibnamefont {Demkowicz-Dobrzanski}},\ }\href {\doibase
  10.1088/1367-2630/15/7/073043} {\bibfield  {journal} {\bibinfo  {journal}
  {New Journal of Physics}\ }\textbf {\bibinfo {volume} {15}},\ \bibinfo
  {pages} {073043} (\bibinfo {year} {2013})} \BibitemShut {NoStop}%
\bibitem [{\citenamefont {Dowling}(2008)}]{Dowling2008}%
  \BibitemOpen
  \bibfield  {author} {\bibinfo {author} {\bibfnamefont {J.~P.}\ \bibnamefont
  {Dowling}},\ }\href {\doibase 10.1080/00107510802091298} {\bibfield
  {journal} {\bibinfo  {journal} {Contemporary Physics}\ }\textbf {\bibinfo
  {volume} {49}},\ \bibinfo {pages} {125} (\bibinfo {year} {2008})}\BibitemShut
  {NoStop}%
\bibitem [{\citenamefont {Israel}\ \emph {et~al.}(2014)\citenamefont {Israel},
  \citenamefont {Rosen},\ and\ \citenamefont {Silberberg}}]{Israel2014}%
  \BibitemOpen
  \bibfield  {author} {\bibinfo {author} {\bibfnamefont {Y.}~\bibnamefont
  {Israel}}, \bibinfo {author} {\bibfnamefont {S.}~\bibnamefont {Rosen}}, \
  and\ \bibinfo {author} {\bibfnamefont {Y.}~\bibnamefont {Silberberg}},\
  }\href {\doibase 10.1103/PhysRevLett.112.103604} {\bibfield  {journal}
  {\bibinfo  {journal} {Physical Review Letters}\ }\textbf {\bibinfo {volume}
  {112}},\ \bibinfo {pages} {12} (\bibinfo {year} {2014})}\BibitemShut
  {NoStop}%
\bibitem [{\citenamefont {Matthews}\ \emph {et~al.}(2016)\citenamefont
  {Matthews}, \citenamefont {Zhou}, \citenamefont {Cable}, \citenamefont
  {Shadbolt}, \citenamefont {Saunders}, \citenamefont {Durkin}, \citenamefont
  {Pryde},\ and\ \citenamefont {O'Brien}}]{Matthews2016}%
  \BibitemOpen
  \bibfield  {author} {\bibinfo {author} {\bibfnamefont {J.~C.}\ \bibnamefont
  {Matthews}}, \bibinfo {author} {\bibfnamefont {X.-Q.}\ \bibnamefont {Zhou}},
  \bibinfo {author} {\bibfnamefont {H.}~\bibnamefont {Cable}}, \bibinfo
  {author} {\bibfnamefont {P.~J.}\ \bibnamefont {Shadbolt}}, \bibinfo {author}
  {\bibfnamefont {D.~J.}\ \bibnamefont {Saunders}}, \bibinfo {author}
  {\bibfnamefont {G.~A.}\ \bibnamefont {Durkin}}, \bibinfo {author}
  {\bibfnamefont {G.~J.}\ \bibnamefont {Pryde}}, \ and\ \bibinfo {author}
  {\bibfnamefont {J.~L.}\ \bibnamefont {O'Brien}},\ }\href {\doibase
  10.1038/npjqi.2016.23} {\bibfield  {journal} {\bibinfo  {journal} {npj
  Quantum Information}\ }\textbf {\bibinfo {volume} {2}},\ \bibinfo {pages}
  {16023} (\bibinfo {year} {2016})}\BibitemShut {NoStop}%
\bibitem [{\citenamefont {Huang}\ \emph {et~al.}(2016)\citenamefont {Huang},
  \citenamefont {Macchiavello},\ and\ \citenamefont {Maccone}}]{Huang2016}%
  \BibitemOpen
  \bibfield  {author} {\bibinfo {author} {\bibfnamefont {Z.}~\bibnamefont
  {Huang}}, \bibinfo {author} {\bibfnamefont {C.}~\bibnamefont {Macchiavello}},
  \ and\ \bibinfo {author} {\bibfnamefont {L.}~\bibnamefont {Maccone}},\ }\href
  {\doibase 10.1103/PhysRevA.94.012101} {\bibfield  {journal} {\bibinfo
  {journal} {Physical Review A}\
  }\textbf {\bibinfo {volume} {94}},\ \bibinfo {pages} {1} (\bibinfo {year}
  {2016})}
  \BibitemShut {NoStop}%
\bibitem [{\citenamefont {Demkowicz-Dobrza{\'{n}}ski}\ and\ \citenamefont
  {Maccone}(2014)}]{Demkowicz-Dobrzanski2014}%
  \BibitemOpen
  \bibfield  {author} {\bibinfo {author} {\bibfnamefont {R.}~\bibnamefont
  {Demkowicz-Dobrza{\'{n}}ski}}\ and\ \bibinfo {author} {\bibfnamefont
  {L.}~\bibnamefont {Maccone}},\ }\href {\doibase
  10.1103/PhysRevLett.113.250801} {\bibfield  {journal} {\bibinfo  {journal}
  {Physical Review Letters}\ }\textbf {\bibinfo {volume} {113}},\ \bibinfo
  {pages} {250801} (\bibinfo {year} {2014})}\BibitemShut {NoStop}%
\bibitem [{\citenamefont {D{\"{u}}r}\ \emph {et~al.}(2014)\citenamefont
  {D{\"{u}}r}, \citenamefont {Skotiniotis}, \citenamefont {Fr{\"{o}}wis},\ and\
  \citenamefont {Kraus}}]{Dur2014}%
  \BibitemOpen
  \bibfield  {author} {\bibinfo {author} {\bibfnamefont {W.}~\bibnamefont
  {D{\"{u}}r}}, \bibinfo {author} {\bibfnamefont {M.}~\bibnamefont
  {Skotiniotis}}, \bibinfo {author} {\bibfnamefont {F.}~\bibnamefont
  {Fr{\"{o}}wis}}, \ and\ \bibinfo {author} {\bibfnamefont {B.}~\bibnamefont
  {Kraus}},\ }\href {\doibase 10.1103/PhysRevLett.112.080801} {\bibfield
  {journal} {\bibinfo  {journal} {Physical Review Letters}\ }\textbf {\bibinfo
  {volume} {112}},\ \bibinfo {pages} {1} (\bibinfo {year} {2014})}\BibitemShut {NoStop}%
\bibitem [{\citenamefont {Jachura}\ \emph {et~al.}(2016)\citenamefont
  {Jachura}, \citenamefont {Chrapkiewicz}, \citenamefont
  {Demkowicz-Dobrza{\'{n}}ski}, \citenamefont {Wasilewski},\ and\ \citenamefont
  {Banaszek}}]{Jachura2016}%
  \BibitemOpen
  \bibfield  {author} {\bibinfo {author} {\bibfnamefont {M.}~\bibnamefont
  {Jachura}}, \bibinfo {author} {\bibfnamefont {R.}~\bibnamefont
  {Chrapkiewicz}}, \bibinfo {author} {\bibfnamefont {R.}~\bibnamefont
  {Demkowicz-Dobrza{\'{n}}ski}}, \bibinfo {author} {\bibfnamefont
  {W.}~\bibnamefont {Wasilewski}}, \ and\ \bibinfo {author} {\bibfnamefont
  {K.}~\bibnamefont {Banaszek}},\ }\href {\doibase 10.1038/ncomms11411}
  {\bibfield  {journal} {\bibinfo  {journal} {Nature Communications}\ }\textbf
  {\bibinfo {volume} {7}},\ \bibinfo {pages} {11411} (\bibinfo {year}
  {2016})} \BibitemShut {NoStop}%
\bibitem [{\citenamefont {Kacprowicz}\ \emph {et~al.}(2010)\citenamefont
  {Kacprowicz}, \citenamefont {Demkowicz-Dobrza{\'{n}}ski}, \citenamefont
  {Wasilewski}, \citenamefont {Banaszek},\ and\ \citenamefont
  {Walmsley}}]{Kacprowicz2010}%
  \BibitemOpen
  \bibfield  {author} {\bibinfo {author} {\bibfnamefont {M.}~\bibnamefont
  {Kacprowicz}}, \bibinfo {author} {\bibfnamefont {R.}~\bibnamefont
  {Demkowicz-Dobrza{\'{n}}ski}}, \bibinfo {author} {\bibfnamefont
  {W.}~\bibnamefont {Wasilewski}}, \bibinfo {author} {\bibfnamefont
  {K.}~\bibnamefont {Banaszek}}, \ and\ \bibinfo {author} {\bibfnamefont
  {I.~A.}\ \bibnamefont {Walmsley}},\ }\href {\doibase 10.1038/nphoton.2010.39}
  {\bibfield  {journal} {\bibinfo  {journal} {Nature Photonics}\ }\textbf
  {\bibinfo {volume} {4}},\ \bibinfo {pages} {357} (\bibinfo {year} {2010})} \BibitemShut
  {NoStop}%
\bibitem [{\citenamefont {Ono}\ \emph {et~al.}(2013)\citenamefont {Ono},
  \citenamefont {Okamoto},\ and\ \citenamefont {Takeuchi}}]{Ono2013}%
  \BibitemOpen
  \bibfield  {author} {\bibinfo {author} {\bibfnamefont {T.}~\bibnamefont
  {Ono}}, \bibinfo {author} {\bibfnamefont {R.}~\bibnamefont {Okamoto}}, \ and\
  \bibinfo {author} {\bibfnamefont {S.}~\bibnamefont {Takeuchi}},\ }\href
  {\doibase 10.1038/ncomms3426} {\bibfield  {journal} {\bibinfo  {journal}
  {Nature Communications}\ }\textbf {\bibinfo {volume} {4}},\ \bibinfo {pages}
  {1} (\bibinfo {year} {2013})}\BibitemShut {NoStop}%
\bibitem [{\citenamefont {{C. K. Hong}}\ \emph {et~al.}(1987)\citenamefont {{C.
  K. Hong}}, \citenamefont {Ou},\ and\ \citenamefont {Mandel}}]{C.K.Hong1987}%
  \BibitemOpen
  \bibfield  {author} {\bibinfo {author} {\bibnamefont {{C. K. Hong}}},
  \bibinfo {author} {\bibfnamefont {Z.}~\bibnamefont {Ou}}, \ and\ \bibinfo
  {author} {\bibfnamefont {L.}~\bibnamefont {Mandel}},\ }\href@noop {}
  {\bibfield  {journal} {\bibinfo  {journal} {Physical Review Letters}\
  }\textbf {\bibinfo {volume} {59}},\ \bibinfo {pages} {2044} (\bibinfo {year}
  {1987})}\BibitemShut {NoStop}%
\bibitem [{\citenamefont {Rarity}\ \emph {et~al.}(1990)\citenamefont {Rarity},
  \citenamefont {Tapster}, \citenamefont {Jakeman}, \citenamefont {Larchuk},
  \citenamefont {Campos}, \citenamefont {Teich},\ and\ \citenamefont
  {Saleh}}]{Rarity1990}%
  \BibitemOpen
  \bibfield  {author} {\bibinfo {author} {\bibfnamefont {J.~G.}\ \bibnamefont
  {Rarity}}, \bibinfo {author} {\bibfnamefont {P.~R.}\ \bibnamefont {Tapster}},
  \bibinfo {author} {\bibfnamefont {E.}~\bibnamefont {Jakeman}}, \bibinfo
  {author} {\bibfnamefont {T.}~\bibnamefont {Larchuk}}, \bibinfo {author}
  {\bibfnamefont {R.~A.}\ \bibnamefont {Campos}}, \bibinfo {author}
  {\bibfnamefont {M.~C.}\ \bibnamefont {Teich}}, \ and\ \bibinfo {author}
  {\bibfnamefont {B.~E.~A.}\ \bibnamefont {Saleh}},\ }\href {\doibase
  10.1103/PhysRevLett.65.1348} {\bibfield  {journal} {\bibinfo  {journal}
  {Physical Review Letters}\ }\textbf {\bibinfo {volume} {65}},\ \bibinfo
  {pages} {1348} (\bibinfo {year} {1990})}\BibitemShut {NoStop}%
\bibitem [{\citenamefont {Demkowicz-Dobrzanski}\ \emph
  {et~al.}(2009)\citenamefont {Demkowicz-Dobrzanski}, \citenamefont {Dorner},
  \citenamefont {Smith}, \citenamefont {Lundeen}, \citenamefont {Wasilewski},
  \citenamefont {Banaszek},\ and\ \citenamefont
  {Walmsley}}]{Demkowicz-Dobrzanski2009}%
  \BibitemOpen
  \bibfield  {author} {\bibinfo {author} {\bibfnamefont {R.}~\bibnamefont
  {Demkowicz-Dobrzanski}}, \bibinfo {author} {\bibfnamefont {U.}~\bibnamefont
  {Dorner}}, \bibinfo {author} {\bibfnamefont {B.~J.}\ \bibnamefont {Smith}},
  \bibinfo {author} {\bibfnamefont {J.~S.}\ \bibnamefont {Lundeen}}, \bibinfo
  {author} {\bibfnamefont {W.}~\bibnamefont {Wasilewski}}, \bibinfo {author}
  {\bibfnamefont {K.}~\bibnamefont {Banaszek}}, \ and\ \bibinfo {author}
  {\bibfnamefont {I.~A.}\ \bibnamefont {Walmsley}},\ }\href {\doibase
  10.1103/PhysRevA.80.013825} {\bibfield  {journal} {\bibinfo  {journal}
  {Physical Review A}\ }\textbf
  {\bibinfo {volume} {80}},\ \bibinfo {pages} {1} (\bibinfo {year} {2009})} \BibitemShut
  {NoStop}%
\bibitem [{\citenamefont {Dorner}\ \emph {et~al.}(2009)\citenamefont {Dorner},
  \citenamefont {Demkowicz-Dobrzanski}, \citenamefont {Smith}, \citenamefont
  {Lundeen}, \citenamefont {Wasilewski}, \citenamefont {Banaszek},\ and\
  \citenamefont {Walmsley}}]{Dorner2009}%
  \BibitemOpen
  \bibfield  {author} {\bibinfo {author} {\bibfnamefont {U.}~\bibnamefont
  {Dorner}}, \bibinfo {author} {\bibfnamefont {R.}~\bibnamefont
  {Demkowicz-Dobrzanski}}, \bibinfo {author} {\bibfnamefont {B.~J.}\
  \bibnamefont {Smith}}, \bibinfo {author} {\bibfnamefont {J.~S.}\ \bibnamefont
  {Lundeen}}, \bibinfo {author} {\bibfnamefont {W.}~\bibnamefont {Wasilewski}},
  \bibinfo {author} {\bibfnamefont {K.}~\bibnamefont {Banaszek}}, \ and\
  \bibinfo {author} {\bibfnamefont {I.~A.}\ \bibnamefont {Walmsley}},\ }\href
  {\doibase 10.1103/PhysRevLett.102.040403} {\bibfield  {journal} {\bibinfo
  {journal} {Physical Review Letters}\ }\textbf {\bibinfo {volume} {102}},\
  \bibinfo {pages} {1} (\bibinfo {year} {2009})} \BibitemShut {NoStop}%
\bibitem [{\citenamefont {MacCone}\ and\ \citenamefont {{De
  Cillis}}(2009)}]{MacCone2009}%
  \BibitemOpen
  \bibfield  {author} {\bibinfo {author} {\bibfnamefont {L.}~\bibnamefont
  {MacCone}}\ and\ \bibinfo {author} {\bibfnamefont {G.}~\bibnamefont {{De
  Cillis}}},\ }\href {\doibase 10.1103/PhysRevA.79.023812} {\bibfield
  {journal} {\bibinfo  {journal} {Physical Review A}\ }\textbf {\bibinfo {volume} {79}},\ \bibinfo {pages} {2}
  (\bibinfo {year} {2009})} \BibitemShut {NoStop}%
\bibitem [{\citenamefont {Escher}\ \emph {et~al.}(2011)\citenamefont {Escher},
  \citenamefont {{de Matos Filho}},\ and\ \citenamefont
  {Davidovich}}]{Escher2011}%
  \BibitemOpen
  \bibfield  {author} {\bibinfo {author} {\bibfnamefont {B.~M.}\ \bibnamefont
  {Escher}}, \bibinfo {author} {\bibfnamefont {R.~L.}\ \bibnamefont {{de Matos
  Filho}}}, \ and\ \bibinfo {author} {\bibfnamefont {L.}~\bibnamefont
  {Davidovich}},\ }\href {\doibase 10.1038/nphys1958} {\bibfield  {journal}
  {\bibinfo  {journal} {Nature Physics}\ }\textbf {\bibinfo {volume} {7}},\
  \bibinfo {pages} {406} (\bibinfo {year} {2011})} \BibitemShut {NoStop}%
\bibitem [{\citenamefont {Pope}\ \emph {et~al.}(2004)\citenamefont {Pope},
  \citenamefont {Wiseman},\ and\ \citenamefont {Langford}}]{Pope2004}%
  \BibitemOpen
  \bibfield  {author} {\bibinfo {author} {\bibfnamefont {D.~T.}\ \bibnamefont
  {Pope}}, \bibinfo {author} {\bibfnamefont {H.~M.}\ \bibnamefont {Wiseman}}, \
  and\ \bibinfo {author} {\bibfnamefont {N.~K.}\ \bibnamefont {Langford}},\
  }\href {\doibase 10.1103/PhysRevA.70.043812} {\bibfield  {journal} {\bibinfo
  {journal} {Physical Review A}\
  }\textbf {\bibinfo {volume} {70}},\ \bibinfo {pages} {1} (\bibinfo {year}
  {2004})}\BibitemShut {NoStop}%
\bibitem [{\citenamefont {Hou}\ \emph {et~al.}(2016)\citenamefont {Hou},
  \citenamefont {Zhu}, \citenamefont {Xiang}, \citenamefont {Li},\ and\
  \citenamefont {Guo}}]{Hou2016}%
  \BibitemOpen
  \bibfield  {author} {\bibinfo {author} {\bibfnamefont {Z.}~\bibnamefont
  {Hou}}, \bibinfo {author} {\bibfnamefont {H.}~\bibnamefont {Zhu}}, \bibinfo
  {author} {\bibfnamefont {G.-Y.}\ \bibnamefont {Xiang}}, \bibinfo {author}
  {\bibfnamefont {C.-F.}\ \bibnamefont {Li}}, \ and\ \bibinfo {author}
  {\bibfnamefont {G.-C.}\ \bibnamefont {Guo}},\ }\href {\doibase
  10.1038/npjqi.2016.1} {\bibfield  {journal} {\bibinfo  {journal} {npj Quantum
  Information}\ }\textbf {\bibinfo {volume} {2}},\ \bibinfo {pages} {16001}
  (\bibinfo {year} {2016})} \BibitemShut {NoStop}%
\bibitem [{\citenamefont {Lerch}\ and\ \citenamefont
  {Stefanov}(2014)}]{Lerch2014}%
  \BibitemOpen
  \bibfield  {author} {\bibinfo {author} {\bibfnamefont {S.}~\bibnamefont
  {Lerch}}\ and\ \bibinfo {author} {\bibfnamefont {A.}~\bibnamefont
  {Stefanov}},\ }\href {\doibase 10.1364/OL.39.005399} {\bibfield  {journal}
  {\bibinfo  {journal} {Optics Letters}\ }\textbf {\bibinfo {volume} {39}},\
  \bibinfo {pages} {5399} (\bibinfo {year} {2014})}\BibitemShut {NoStop}%
\bibitem [{\citenamefont {Okamoto}\ \emph {et~al.}(2012)\citenamefont
  {Okamoto}, \citenamefont {Iefuji}, \citenamefont {Oyama}, \citenamefont
  {Yamagata}, \citenamefont {Imai}, \citenamefont {Fujiwara},\ and\
  \citenamefont {Takeuchi}}]{Okamoto2012}%
  \BibitemOpen
  \bibfield  {author} {\bibinfo {author} {\bibfnamefont {R.}~\bibnamefont
  {Okamoto}}, \bibinfo {author} {\bibfnamefont {M.}~\bibnamefont {Iefuji}},
  \bibinfo {author} {\bibfnamefont {S.}~\bibnamefont {Oyama}}, \bibinfo
  {author} {\bibfnamefont {K.}~\bibnamefont {Yamagata}}, \bibinfo {author}
  {\bibfnamefont {H.}~\bibnamefont {Imai}}, \bibinfo {author} {\bibfnamefont
  {A.}~\bibnamefont {Fujiwara}}, \ and\ \bibinfo {author} {\bibfnamefont
  {S.}~\bibnamefont {Takeuchi}},\ }\href {\doibase
  10.1103/PhysRevLett.109.130404} {\bibfield  {journal} {\bibinfo  {journal}
  {Physical Review Letters}\ }\textbf {\bibinfo {volume} {109}},\ \bibinfo
  {pages} {1} (\bibinfo {year} {2012})} \BibitemShut {NoStop}%
\bibitem [{\citenamefont {Higgins}\ \emph {et~al.}(2007)\citenamefont
  {Higgins}, \citenamefont {Berry}, \citenamefont {Bartlett}, \citenamefont
  {Wiseman},\ and\ \citenamefont {Pryde}}]{Higgins2007}%
  \BibitemOpen
  \bibfield  {author} {\bibinfo {author} {\bibfnamefont {B.~L.}\ \bibnamefont
  {Higgins}}, \bibinfo {author} {\bibfnamefont {D.~W.}\ \bibnamefont {Berry}},
  \bibinfo {author} {\bibfnamefont {S.~D.}\ \bibnamefont {Bartlett}}, \bibinfo
  {author} {\bibfnamefont {H.~M.}\ \bibnamefont {Wiseman}}, \ and\ \bibinfo
  {author} {\bibfnamefont {G.~J.}\ \bibnamefont {Pryde}},\ }\href {\doibase
  10.1038/nature06257} {\bibfield  {journal} {\bibinfo  {journal} {Nature}\
  }\textbf {\bibinfo {volume} {450}},\ \bibinfo {pages} {393} (\bibinfo {year}
  {2007})}\BibitemShut {NoStop}%
\bibitem [{\citenamefont {Chapeau-Blondeau}(2016)}]{Chapeau-Blondeau2016}%
  \BibitemOpen
  \bibfield  {author} {\bibinfo {author} {\bibfnamefont {F.}~\bibnamefont
  {Chapeau-Blondeau}},\ }\href {\doibase 10.1103/PhysRevA.94.022334} {\bibfield
   {journal} {\bibinfo  {journal} {Physical Review A}\ }\textbf {\bibinfo
  {volume} {94}},\ \bibinfo {pages} {022334} (\bibinfo {year}
  {2016})}\BibitemShut {NoStop}%
\bibitem [{\citenamefont {Nielsen}\ and\ \citenamefont
  {Chuang}(2011)}]{Nielsen2011}%
  \BibitemOpen
  \bibfield  {author} {\bibinfo {author} {\bibfnamefont {M.~A.}\ \bibnamefont
  {Nielsen}}\ and\ \bibinfo {author} {\bibfnamefont {I.~L.}\ \bibnamefont
  {Chuang}},\ }\href {\doibase 10.1017/CBO9780511976667} {\emph {\bibinfo
  {Quantum Computation and Quantum Information} {Cambridge University Press}}}\ (\bibinfo {year} {2011}) \BibitemShut {NoStop}%
  

\end{thebibliography}
\end{document}